\documentstyle[12pt,amstex,amssymb,aps]{revtex}

\input BoxedEPS
\SetRokickiEPSFSpecial  
\HideDisplacementBoxes

\draft

\tighten

\begin{document}
\title{Instability of a Bose-Einstein Condensate\\ with Attractive
Interaction}
\author{Antonios Eleftheriou and Kerson Huang}
\address{Department of Physics and Center for Theoretical Physics \\
Massachusetts Institute of Technology\\
Cambridge, MA 02139, USA}
\date{\today}
\maketitle
\pacs{03.75.Fi, 42.65.Jx, 32.80.Pj\hfil MIT-CTP \#2886}

\begin{abstract}
We study the stability of a Bose-Einstein condensate of harmonically
trapped atoms with negative scattering length, specifically
$^{7}$Li. Our method is to solve the time-dependent nonlinear
Schr\"{o}dinger equation numerically.  For an isolated condensate,
with no gain or loss, we find that the system is stable (apart from
quantum tunneling) if the particle number $N$ is less than a critical
number $N_{c}.$ For $N>N_{c}$, the system collapses to high-density
clumps in a region near the center of the trap. The time for the onset
of collapse is on the order of 1 trap period. Within numerical
uncertainty, the results are consistent with the formation of a
``black hole'' of infinite density fluctuations, as predicted by Ueda
and Huang \cite{UedaHuang}. We obtain numerically $N_{c}\approx
1251$. We then include gain-loss mechanisms, i.e., the gain of atoms
from a surrounding ``thermal cloud'', and the loss due to two- and
three-body collisions. The number $N$ now oscillates in a steady
state, with a period of about 145 trap periods. We obtain $N_{c}\approx
1260$ as the maximum value in the oscillations.
\end{abstract}

\section{Introduction and Summary}

Bose-Einstein condensation has been observed in magnetically trapped dilute
vapors of$\ $\ the alkali elements $^{87}$Rb~\cite{Anderson}, $^{23}$Na~\cite
{Davis}, $^{7}$Li~\cite{Bradley}, and $^{1}$H~\cite{Fried}. At the
nanodegree temperatures of these experiments, the systems would have frozen
solid long ago were they in free space. In the confining trap, however,
zero-point motion keeps the atoms apart, and the systems remain gaseous. The
case of $^{7}$Li is special, however, in that the interatomic interaction is
predominantly attractive, as indicated by a negative scattering length.
Thus, the condensate in $^{7}$Li should be less stable than the other cases.
The purpose of this paper is to study the nature of the instability, its
onset, and manifestations.

The object of study is the condensate wave function $\psi ({\bf r},t)$,
which gives the probability amplitude for annihilating one particle in the
condensate at ${\bf r}$ at time $t$. We use the time-dependent
Gross-Pitaevskii (GP) equation, or nonlinear Schr\"{o}dinger equation (NLSE),
which corresponds to a mean-field approximation:

\begin{eqnarray}
i\hbar \frac{\partial \psi }{\partial t} &=&\left[ -\frac{\hbar ^{2}}{2m}%
\nabla ^{2}+V(r)-U_{0}|\psi |^{2}\right] \psi  \nonumber \\
U_{0} &=&\frac{4\pi \hbar ^{2}|a|}{m}  \label{NLSE}
\end{eqnarray}
where $a$ is a negative scattering length, and the external potential is
taken to be harmonic: 
\begin{equation}
V(r)=\frac{1}{2}m\omega ^{2}r^{2}
\end{equation}
This defines a characteristic length $d_{0}$, the width of the unperturbed
ground-state wave function: 
\begin{equation}
d_{0}=\sqrt{\frac{\hbar }{m\omega }}
\end{equation}
The number of condensate particles enters through the normalization 
\begin{equation}
N=\int d^{3}r|\psi |^{2}
\end{equation}
which is a constant of the motion. Another constant of the motion is the
Hamiltonian 
\begin{equation}
H=\int d^{3}r\left[ -\frac{\hbar ^{2}}{2m}\psi ^{\ast }\nabla ^{2}\psi
+V(r)\psi ^{\ast }\psi -\frac{U_{0}}{2}(\psi ^{\ast }\psi )^{2}\right]
\label{Hamiltonian}
\end{equation}
The parameters used in $^{7}$Li~ experiments correspond to \cite{Bradley} 
\begin{align}
a& =-1.45\text{ nm}  \nonumber \\
d_{0}& \approx \text{3.16 }\mu \text{m}  \nonumber \\
\omega & \approx 908\text{ s}^{-1}  \label{parameters}
\end{align}
where $\omega$ is taken to be approximately equal to the geometric
mean of the three circular frequencies of the experimental
trap. Equation (\ref{NLSE}) does not \ take into account the coupling
between the condensate and the ``thermal cloud'' of uncondensed atoms,
nor does it describe the loss of atoms from the trap\ due to
collisions. These effects will be considered later.

\ In free space, the NLSE with attractive interactions has an instability
known in plasma physics as\ ``self-focusing'' \cite{Zakharov75}, whereby the
initial wave function develops a singularity in finite time, corresponding
to a local collapse to a state of infinite density. In an external trap,
however, the ground state is apparently stable, as long as $N$ is not too
large, as indicated by variational calculations using a Gaussian trial wave
function \cite{Baym}. The width of the Gaussian narrows with increasing $N$,
and collapses to zero at some critical value $N_{c}.$ This conclusion is
borne out by other studies \cite{Ruprecht,Dalfovo,Ueda,Kim,Wadati}. In
particular, Kim and Zubarev~obtain an exact upper bound for $N_{c}$, which
for a Gaussian wave function gives 
\begin{equation}
N_{c}\leqq 0.671\,\frac{d_{0}}{|a|}
\end{equation}
Ueda and Leggett \cite{Ueda} obtain the upper bound in a variational
calculation. For the $^{7}$Li~ parameters (\ref{parameters}), this formula
gives $N_{c}\leqq 1463$. In the numerical calculations described later, we
obtain 
\begin{equation}
N_{c}=0.574\frac{d_{0}}{|a|}  \label{Nc}
\end{equation}
or $N_{c}=1251$ for the $^{7}$Li parameters. This number becomes slightly
larger when gain and loss effects are taken into account.

Actually, even ignoring collisional loss, the condensate is only
metastable for $N<N_{c}$, for it can decay via quantum tunneling. This
effect is not described by the NLSE, although an approximate decay
amplitude can be obtained by continuation of the NLSE to imaginary
time. Kagan et al.~ \cite {Kagan} estimated the amplitude by
calculating an overlap integral between initial and final states
represented by Gaussian wave functions, and obtained $(d_{{\rm
f}}/d_{{\rm i}})^{3N/2}$, where $d_{{\rm i}}$ and $d_{{\rm f}}$ are
respectively the widths of the initial and final wave
functions. Through numerical calculations, Shuryak~ \cite{Shuryak}
found that the decay rate is proportional to $\exp [-{\rm
const.}(N_{{\rm c}}-N)]$. Stoof~ \cite{Stoof} wrote down a WKB formula
for the decay rate, but did not explicitly evaluate it. Using
variational wave functions, Ueda and Leggett ~\cite{Ueda} obtained a
rate proportional to $\exp [-{\rm const.}(N_{{\rm c}}-N)^{5/4}].$
These estimates indicate that the tunneling probability is negligible
unless $N\approx N_{\text{{\rm c}}}$. However, the WKB approximation
breaks down in this neighborhood, and reliable calculations become
difficult. In experimental terms, it is also difficult to observe
tunneling in the present system, because the effect is masked by
collisional loss, especially near $N=N_{\text{{\rm c}}}$
\cite{Ketterle}. For these\ reasons, we shall not calculate the
tunneling amplitude in this paper; but we will give a qualitative
discussion of \ the phenomenon, in as far as it pertains to the
instability of the system.

Ueda and Huang \cite{UedaHuang} formulated an approach to the
stability problem, including tunneling, in terms of the Feynman path
integral for a transition amplitude. In a saddle-point approximation
to the path integral, one obtains the NLSE as the saddle-point
condition, whose continuation into imaginary time yields the tunneling
amplitude. Starting with a Gaussian initial wave function, they assume
that it remains Gaussian, in order to analyze the problem
analytically. In this ``Gaussian approximation'', it was found that,
as soon as $N>$ $N_{c}$, the system begins to collapse locally, in a
region at the center of the trap. The size of this region grows as $N$
increases. By putting the NLSE in hydrodynamic form, one can see that
the pressure becomes negative inside this ``black hole'', and,
consequently, there would be infinite density fluctuations.

In the present work, we remove the Gaussian approximation by
performing numerical calculations. We verify that the picture
presented by Ueda and Huang is correct. For $N>$ $N_{c}$ , a
``black-hole'' does appear at the center of the trap, within which the
density has sharp local maxima, whose heights grow as the grid size
for the numerical computation is decreased. This is consistent with
the expectation that the density becomes divergent in the continuum
limit. Finally, we take into account gain and loss mechanisms by
adding phenomenological terms to the NLSE. We are able to study in
some detail the growth-collapse cycles of the condensate, as
anticipated in earlier work \cite{Hulet2,Kagan2}.

\section{Tunneling and Collapse}

To give an overview of the mechanisms for instability, we think of the
condensate wave function at each spatial point as a ``coordinate,''
which moves in an effective potential, consisting of the last two
terms in the Hamiltonian (\ref{Hamiltonian}), plus fluctuation
corrections arising from the term $\psi ^{\ast }\nabla ^{2}\psi $. The
system is classically stable when this motion is bounded at all
spatial points. We immediately see that the attractive interaction
term $-\frac{U_{0}}{2}(\psi ^{\ast }\psi )^{2}$ can lead to
instability, for it decreases without bound when the density
increases. The question is whether there is an energy barrier guarding
against the free fall.

\begin{figure}
$$
\BoxedEPSF{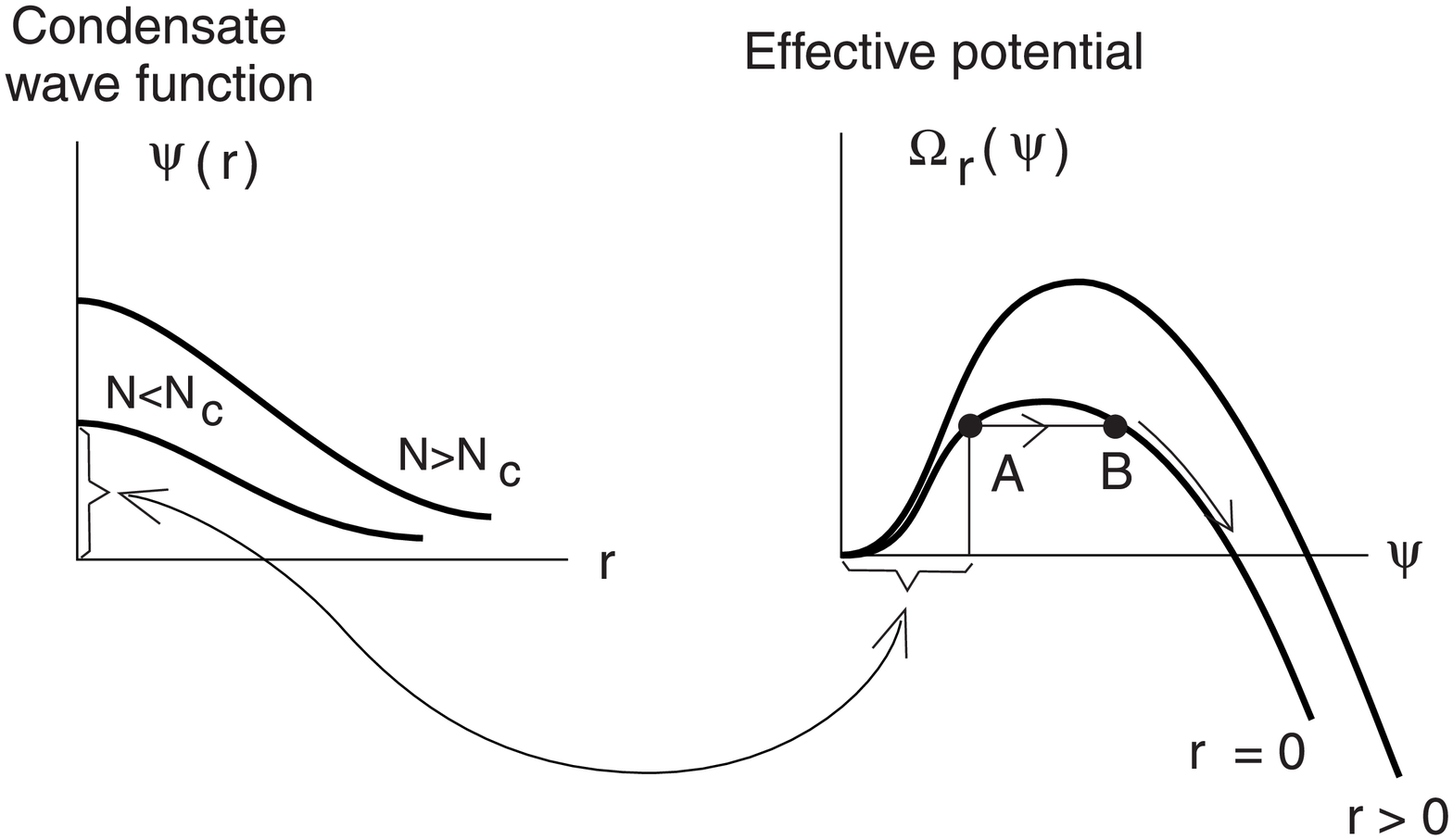 scaled 500}  
$$
\caption{Transfer a local value of the wave function from the left panel to
the horizontal axis on the right. If it lands to the left of the potential
barrier, the local system is classically stable, but can decay via
tunneling, as indicated by the path $A\rightarrow B$. If it lands to the
right, the local system will rapidly collapse towards a state of infinite
density.}
\label{AEKHFig1}
\end{figure}

A precise definition for the effective potential can be modeled after that
for a spatially uniform system in quantum field theory\cite{Qft}. For a
qualitative description, however, it is much simpler to use the ``Gaussian
approximation''$\ $introduced by Ueda and Huang\cite{UedaHuang}, where we
assume that the wave function is Gaussian, given by its initial form 
\begin{equation}
\psi _{0}(r)=C_{0}\exp \left( -r^{2}/d^{2}\right)
\end{equation}
where 
\begin{equation}
C_{0}=\pi ^{-3/4}d^{-3/2}\sqrt{N}
\end{equation}
We introduce a width parameter $\alpha $ by putting 
\begin{equation}
d=\frac{d_{0}}{\sqrt{\alpha }}
\end{equation}
Thus, $\alpha =1$ corresponds to the ground-state eigenfunction of the
harmonic oscillator. With this, we have 
\begin{equation}
-\frac{\hbar ^{2}}{2m}\nabla ^{2}\psi _{0}=\chi (r)\psi _{0}
\end{equation}
where 
\begin{equation}
\chi (r)=\frac{\hbar ^{2}}{2md_{0}^{2}}\left[ 3\alpha -\left( \frac{r}{d_{0}}%
\right) ^{2}\alpha ^{2}\right]
\end{equation}
and in this approximation the effective potential is 
\begin{equation}
\Omega _{r}(\psi )=\left[ V(r)+\chi (r)\right] |\psi |^{2}-\frac{U_{0}}{2}%
|\psi |^{4}
\end{equation}
This is depicted on the right panel of Fig.~\ref{AEKHFig1} for different $r$. There is an
energy barrier with height 
\begin{equation}
W_{r}=\frac{1}{U_{0}}\left[ V(r)+\chi (r)\right] ^{2}
\end{equation}
which is nonzero at $r=0$, because $\chi (0)\neq 0$. At large $|\psi |$ the
effective potential tends to $-\infty $. In reality, of course, the NLSE
ceases to be valid somewhere along the drop, for other physical effects, such
as solidification, come in.

A typical initial wave function $\psi _{0}(r)$ is shown on the left panel of
Fig.~\ref{AEKHFig1}. At a given $r$, we can measure the wave function on the left
panel, and transfer it to the horizontal axis on the right panel. If it lands to
the left of the barrier maximum for the particular $r,$ then the system at
that point is classically stable, but can decay via quantum tunneling as
indicated by the classically forbidden path $A\rightarrow B$. If it lands to
the right, the system at that $r$ will rapidly collapse to a state of high
density.

Since $\psi _{0}(r)$ has a maximum at $r=0$ with value proportional to $%
\sqrt{N}$, the system may only decay via tunneling at any $r$, if $\ $it is
classically stable at $r=0$. Otherwise, the system in a region about $r=0$
will rapidly collapse. The condition for stability against collapse is
therefore 
\begin{equation}
\Omega _{0}(\psi _{0}(0))<W_{0}
\end{equation}
which corresponds to $N<N_{c}$, with 
\begin{equation}
N_{c}=\frac{3}{8}\sqrt{\frac{\pi }{\alpha }}\frac{d_{0}}{|a|}=\frac{0.665}{%
\sqrt{\alpha }}\frac{d_{0}}{|a|}  \label{NcGauss}
\end{equation}
The time for the onset of \ local collapse is expected to be greater than
the time needed to establish a quasi-stationary initial wave function. Thus, 
$\alpha $ should correspond to the best Gaussian approximation to that wave
function. This expectation is verified by numerical calculations described
below, which also show that the Gaussian approximation is very good, up to
the onset of collapse. Of course, $\alpha$ ceases to have meaning
once the collapse begins, for $N>N_{c}$. The numerical calculations
for $N=1248$, which is just below $N_{c}$, give 
\begin{equation}
\alpha =1.79  \label{alpha}
\end{equation}

As mentioned earlier, we will not calculate the tunneling amplitude in this
paper, but concentrate on numerical solutions of the NLSE in real time.

\section{Stationary State}

We seek spherically symmetric solutions to the NLSE (\ref{NLSE}), for the
parameters (\ref{parameters}) for the $^{7}$Li experiments. For convenience
we put 
\begin{equation}
\psi (r,t)=\sqrt{\frac{N}{4\pi d_{0}}}\frac{u(r,t)}{r}
\end{equation}
with the boundary condition $u(0)=0$, and the normalization condition 
\begin{equation}
\frac{1}{d_{0}}\int_{0}^{\infty }dr|u(r,t)|^{2}=1
\end{equation}
From now on, we shall measure distance in units of $d_{0}$, and time in
units of $2\omega ^{-1}$. To do this without introducing new symbols for
dimensionless distance and time, we formally put $d_{0}=\omega /2=1$. The
NLSE (\ref{NLSE}) then takes the form 
\begin{equation}
i\frac{\partial u}{\partial t}=\left( -\frac{\partial ^{2}}{\partial r^{2}}%
+r^{2}-\frac{2G|u|^{2}}{r^{2}}\right) u
\end{equation}
where 
\begin{equation}
G=\frac{N|a|}{d_{0}}=\frac{N}{2180}
\end{equation}
The initial condition with Gaussian form is 
\begin{equation}
u(r,0)=u_{0}r\exp \left( -\frac{1}{2}\alpha r^{2}\right) 
\end{equation}
where 
\begin{equation}
u_{0}=2\pi ^{-1/4}\alpha ^{3/4}
\end{equation}
We calculate $u$ numerically as a complex function, using an implicit
difference equation algorithm due to Goldberg et al.\cite{Schwartz}.

\begin{figure}
$$
\BoxedEPSF{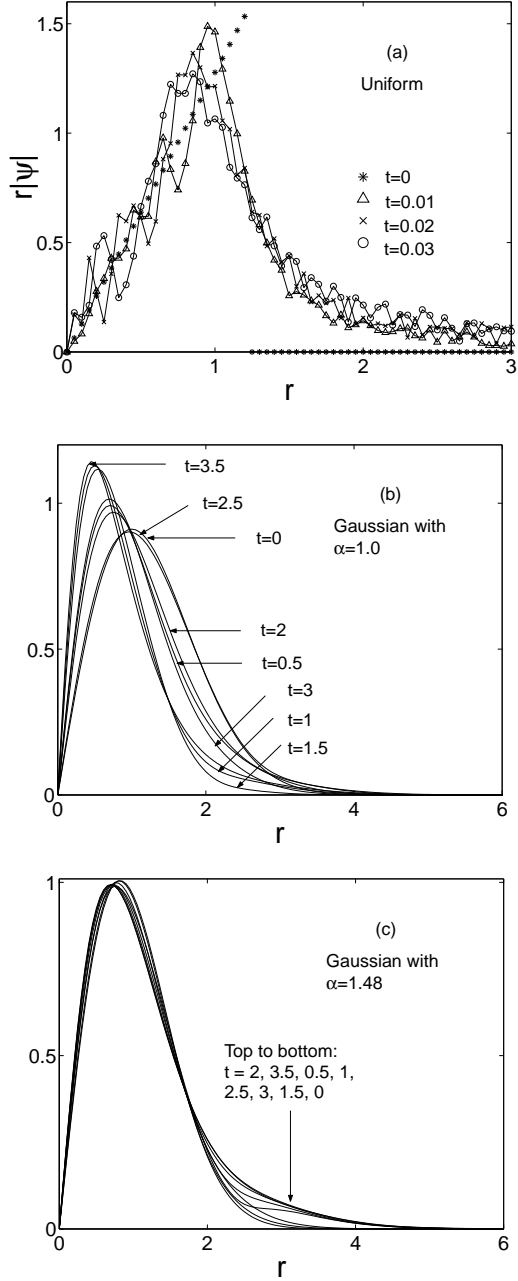 scaled 750}  
$$
\caption{Time development of the condensate wave function. The time
steps are given in units of $2\omega ^{-1}$, where $\omega $ is the
circular frequency of the external harmonic trap. The number of
particles is $N=1142$ (i.e. $G=N/2180=0.524$), which is the highest
particle number for which a Gaussian initial wave function with
$\alpha=1.0$ is stable. (a) An initially uniform wave function tends
to Gaussian form, except for local fluctuations, after a few time
steps. (b) For an initial unperturbed Gaussian wave function with
width $d_{0}$, the width narrows to a stationary value after a few
time steps. (c) Starting from a Gaussian with width
$d_{0}/\sqrt{1.48}$ gives a sensibly stationary wave function.
}
\label{AEKHFig2}
\end{figure}

To find out how various initial states relax to a stationary state, we
first solve the equation with an initial wave function uniform within
a certain radius, and zero outside. For $N=1142$, which is somewhat
below the critical number, the time development is shown in
Fig.~\ref{AEKHFig2}(a). We see that the modulus $r|\psi|$ appears to be
Gaussian-like, except for small local fluctuations, after only about
one-hundredth of an oscillator cycle. In Fig.~\ref{AEKHFig2}(b), we start with a
Gaussian with $\alpha =1$, with the same $N$. This corresponds to the
stationary wave function in the absence of interatomic interaction. In
a few oscillator cycles, the width narrows to a stationary value. If
we initially choose $\alpha $ near the stationary value, the location
of the peak of $r|\psi|$ will slowly oscillate about some mean
position. Fig.~\ref{AEKHFig2}(c) shows the case when the ratio of the standard
deviation to the mean of the peak location is minimized. This
corresponds to $\alpha =1.48$, for this choice of $N$. The wave
function in this case is sensibly stationary, and the Gaussian
approximation is good.

\begin{figure}
$$
\BoxedEPSF{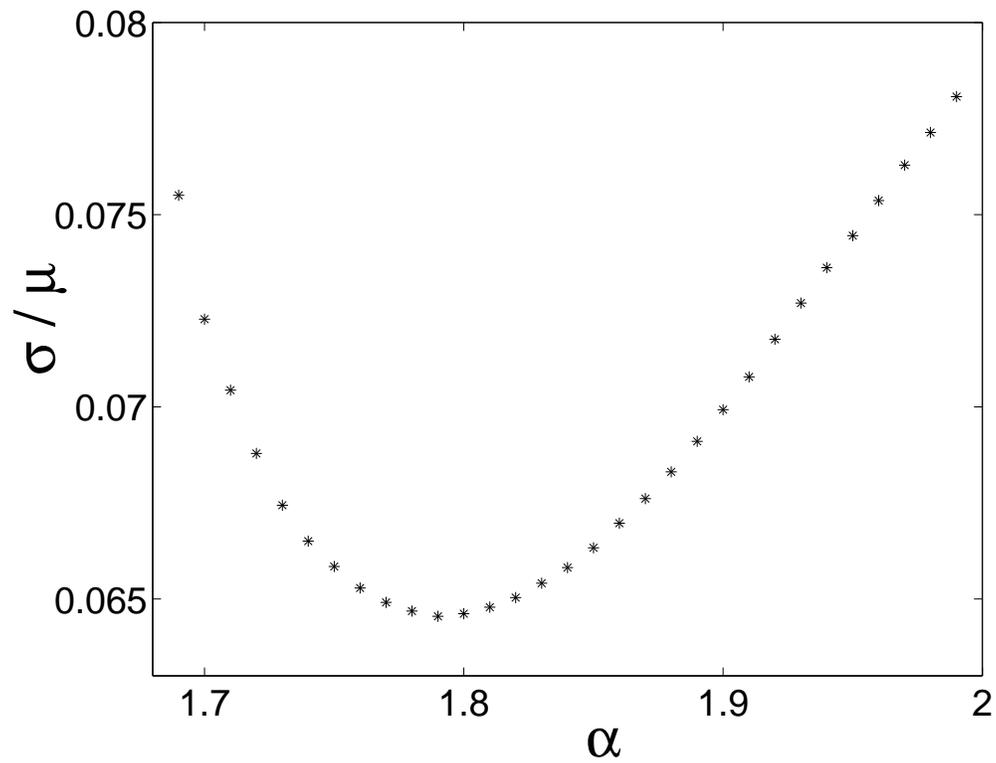 scaled 700}  
$$
\caption{Determining the best width $d=d_{0}/\sqrt{\alpha }$ of the
initial Gaussian wave function $\psi _{0}(r)$, for $N=1248$ ($G=0.5725$), by
minimizing the ratio of the standard deviation of the peak location to
the mean peak location  of $|u(r,t)|$. The minimum occurs at $\alpha =1.79$.}
\label{AEKHFig3}
\end{figure}

In Fig.~\ref{AEKHFig3}, we plot the ratio of the standard deviation to the mean of
the peak location for $N=1248$ (just below critical) against $\alpha$,
and exhibit the minimum at $\alpha =1.79$, as quoted earlier in (\ref
{alpha}). For a quantitative measure of how good the Gaussian
approximation is, we calculate the absolute value of the overlap
between $u(r,t)$ and $u(r,0)$ and find that it lies between 1 and
0.9795, with an average of 0.9906, for $t$ between 0 and 50 ($t$ in
units of 2${\omega}^{-1}$).

\section{Local Collapse}

\begin{figure}
$$
\BoxedEPSF{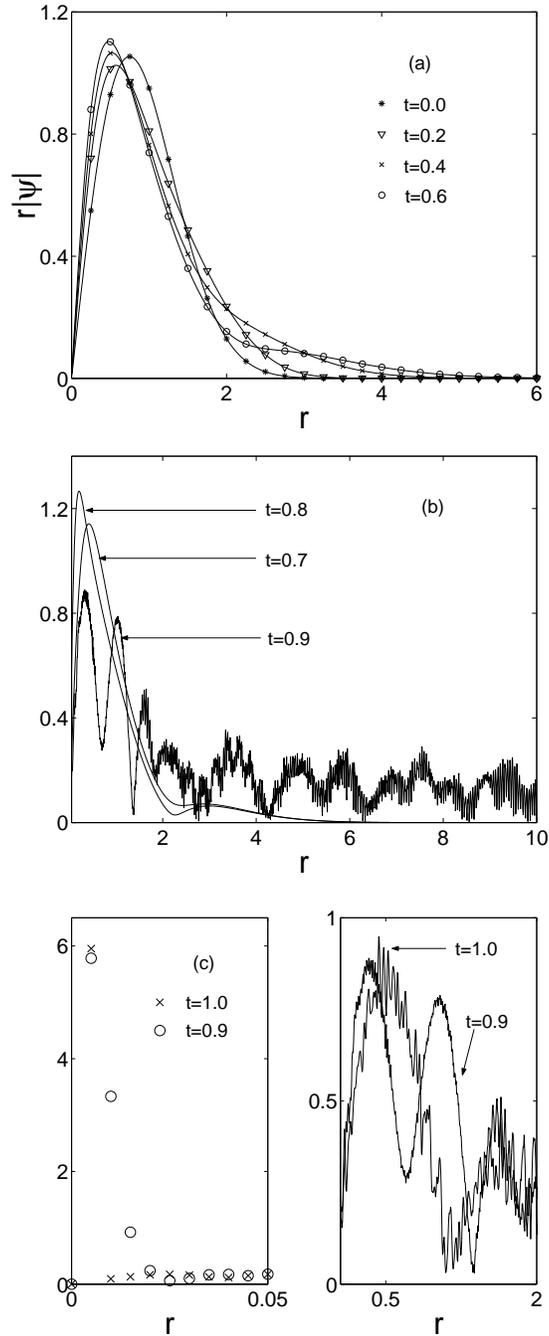 scaled 785}  
$$
\caption{Time development of unstable case, with $N=1380$. (a) The wave function
begins to narrow after
$t
\approx 0.2$. (b) The onset of collapse is signaled by the first occurrence of a dip
(local minimum) in the wave function within a small distance from the
origin, which indicates that a shell of particles begins to
implode. The dip first appears at $t \approx 0.7$ and moves quickly
towards $r=0$. After the onset, the collapse continues by drawing in
particles from all distances, in ripples of implosive flow. In this
figure the first 10 grid points ($r \leqq 0.05$) were cut off for
clarity. (c) The first local minimum of the wave function is now at
less than $r=0.05$. The wave function becomes more ragged as time goes
on. Because of the finiteness of the grid the density at the center
remains finite.}
\label{AEKHFig4}
\end{figure}

To find $N_{c}$, we examine the time development for different $N$, to
look for the onset of local collapse. The signature of the collapse is
the occurrence of a minimum in the wave function within a small
distance from the origin as shown in Fig.~\ref{AEKHFig4} for $\alpha =1.79$ and
$N=1308$. The initial wave function is stationary for a while, but
begins to narrow after $t \approx 0.2$ (Fig.~\ref{AEKHFig4}(a)), and a dip occurs at
$t\approx 0.7$ (Fig.~\ref{AEKHFig4}(b)), at $r\approx 2.3$, which describes the
implosion of a shell of particles at that radius. The first local
minimum moves inwards with time and it eventually gets very close to
$r=0$, at $t \approx 0.9$, as shown in Fig.~\ref{AEKHFig4}(c). The density in the
inner region can grow only by taking particles from the outside, since
the number of particles is conserved. An examination of the phase of
the wave function, whose gradient gives the superfluid velocity, shows
that the superfluid velocity has a peak at the dip, and is directed
inward, as we expect from the continuity equation. Thus,
qualitatively, the onset of collapse happens on a time scale an order
of magnitude greater than that for the establishment of the initial
wave function, and it is initiated by a sudden inward rush of
particles from a finite distance from the center. This signals the
formation of a ``black hole'', as suggested by Ueda and Huang
\cite{UedaHuang}.

\begin{figure}
$$
\BoxedEPSF{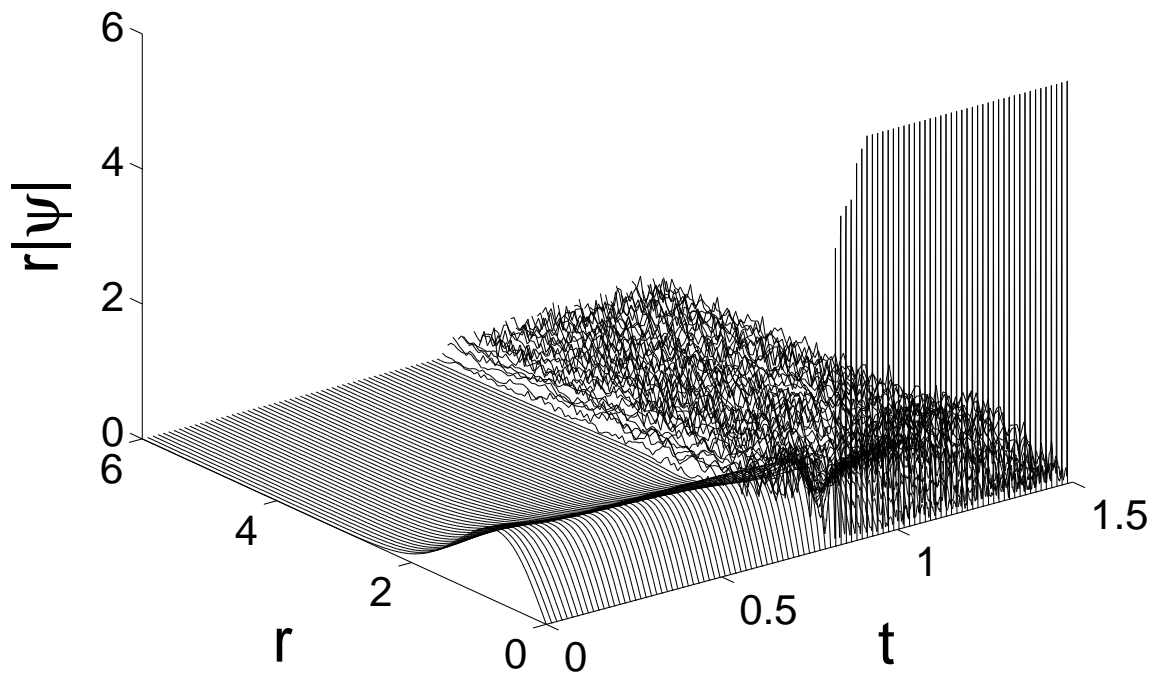 scaled 785}  
$$
\caption{Space-time behavior of the wave function of Fig.~\protect\ref{AEKHFig4} in a
3D plot. The black hole formation is indicated by a sudden emergence of the high
plateau. After that time, the terrain in the plain below the plateau becomes rough,
indicating the ripples of implosive flow. As the spatial grid size tends to zero, the
height of the plateau seems to increase without limit. (See Fig.~\protect\ref{AEKHFig8}.)}
\label{AEKHFig5}
\end{figure}

\begin{figure}
$$
\BoxedEPSF{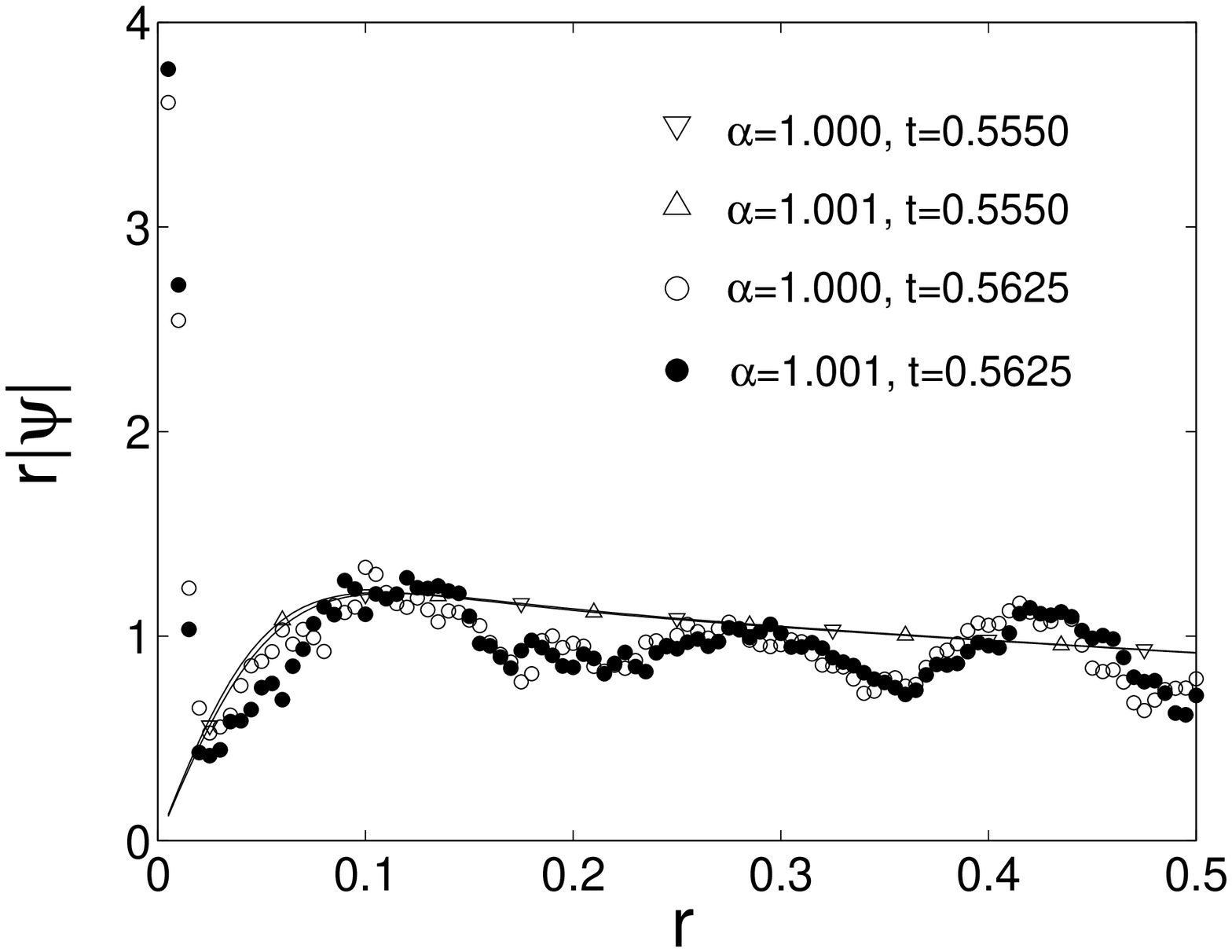 scaled 700} 
$$
\caption{Sensitivity of ripples to the initial conditions. We show the
amplitude of the wave function just before and after the onset of the
collapse for two nearly identical initial wave functions, Gaussians
with $\alpha=1$ and $\alpha=1.001$. Notice that, before the collapse,
the two wave functions practically coincide. However, once the collapse
occurs, they diverge very quickly.}
\label{AEKHFig6}
\end{figure}

After the onset of collapse, the density in the black hole increases
rapidly, as indicated in Fig.~\ref{AEKHFig4}(c). The increase is fueled by particles
drawn from throughout the outer region, but the density outside does
not decrease uniformly. Instead, there are ripples of small implosions
This is illustrated in the 3D plot of Fig.~\ref{AEKHFig5}, with $r|\psi|$ plotted
above the $r$-$t $ plane. The black hole formation is indicated by the
sudden rise of $r|\psi|$ near the origin. The terrain is smooth before the
rise, but very rough after that. The uneven bumps in the terrain are
of course deterministic, but they are very sensitive to variations in
the initial wave function, and in this sense they are random (see
Fig.~\ref{AEKHFig6}). The fact that the height of the plateau seems to remain
constant in time is an artifact of the finite spatial grid size. The
height appears to increase without bound in the continuum limit, as
illustrated in Fig.~\ref{AEKHFig8}.

\begin{figure}
$$
\BoxedEPSF{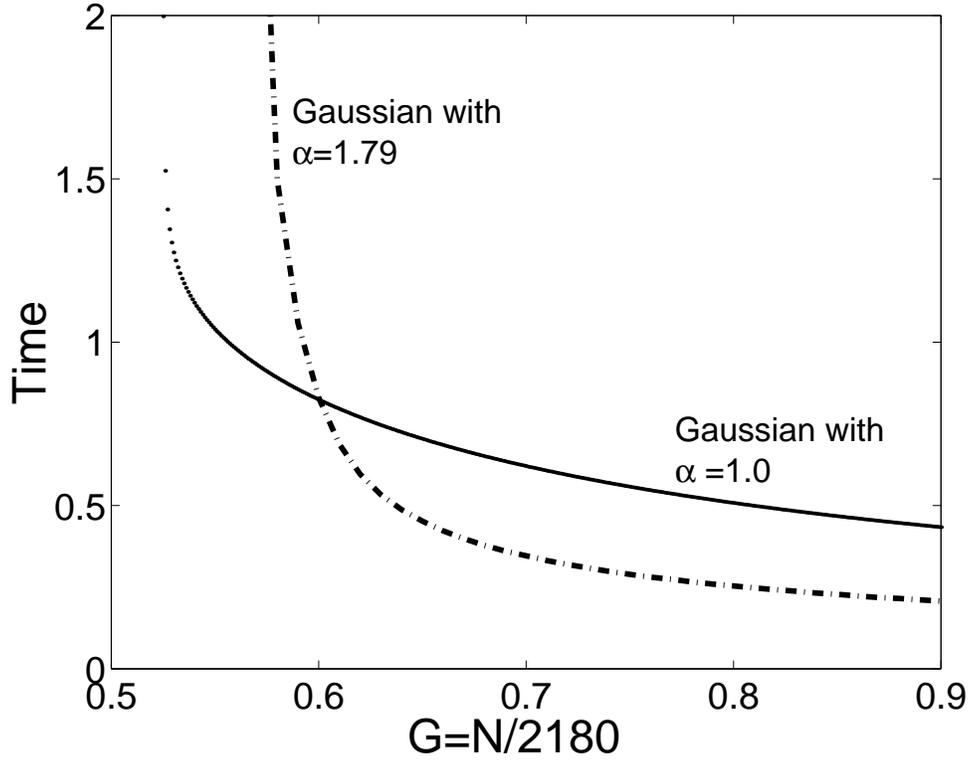 scaled 700} 
$$
\caption{Time for the onset of collapse {\it vs} $N/2180$. The two curves
correspond to a Gaussian initial wave function with $\alpha=1.0$ and to
a Gaussian initial wave function with $\alpha=1.79$. The critical
number $N_{c}$ is determined by the asymptote at which the time goes
to infinity. We get $N_{c}=1145$ ($G_{c}=00.525$) when $\alpha=1.0$ and
$N_{c}=1251$ ($G_{c}=0.574$) when $\alpha=1.79$. The results for the
critical number are not very sensitive to the initial wave function.
}
\label{AEKHFig7}
\end{figure}

\begin{figure}
$$
\BoxedEPSF{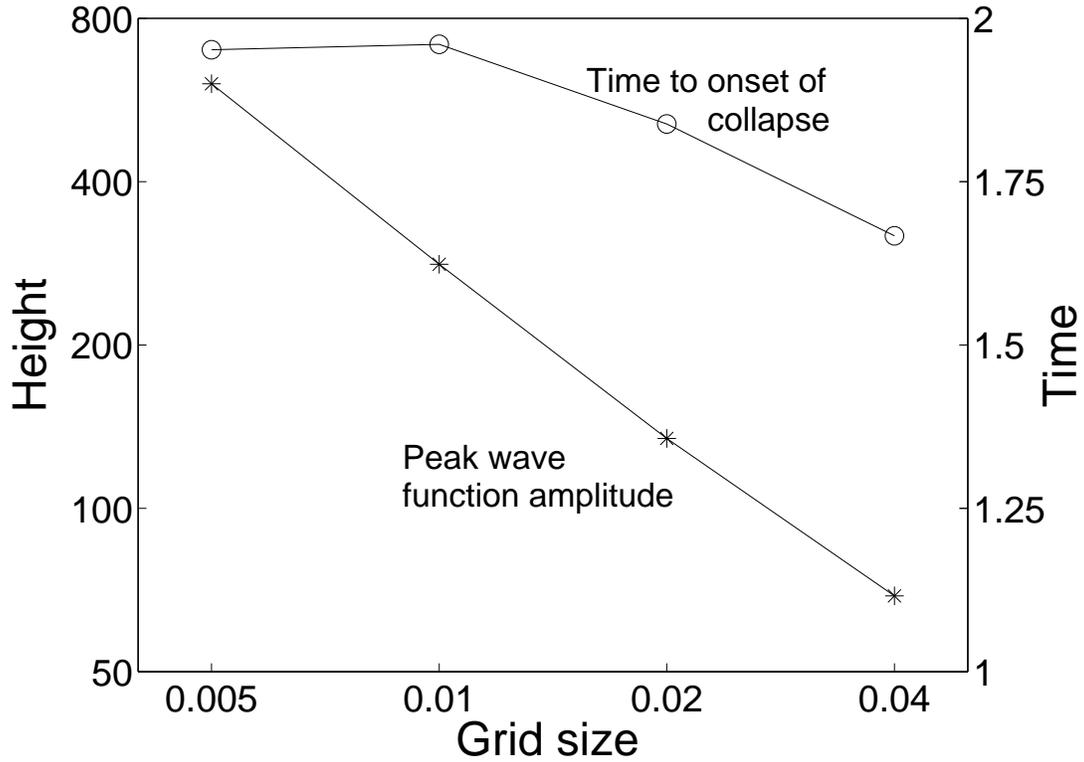 scaled 700} 
$$
\caption{Sensitivity of results on the spatial grid size, measured in
units of $d_{0}$.  The initial wave function is Gaussian with
$\alpha=1.0$.  Circles represent the time for the onset of collapse,
which tends to a finite value in the continuum limit. Asterisks
indicate the height of the peak of the square root of the density in
the collapsed region. It appears to increase without limit as the grid
size tends to zero. With the log-log axes used, the best linear fit
has a slope of slightly less than $-1$, which implies that the height of
the peak increases faster than (grid size)$^{-1}$ as the grid size
goes to zero.}
\label{AEKHFig8}
\end{figure}

To determine $N_{c}$ more precisely, we plot the occurrence time of
the first local minimum in the wave function within distance $r=0.05$
from the origin, as a function of $G=N/2180$. This is done in Fig.~\ref{AEKHFig7},
for a Gaussian initial wave function with $\alpha =1.0$, and for an optimal
Gaussian initial wave function with $\alpha =1.79$.  The results are as
follows:
\begin{equation}
N_{c}=\left\{ 
\begin{array}{lll}
1145 & \text{(Gaussian initial state with}& \alpha = 1.00\text{)} \\
1251 & \text{(Gaussian initial state with}& \alpha = 1.79\text{)} 
\end{array}
\right. 
\end{equation}
which conforms to our expectation that the numbers are not very sensitive to
initial conditions, because the time for the onset of collapse is long
compared to the formation time of a quasi-stationary wave function.

In Fig.~\ref{AEKHFig8} we show the dependence of some of our results on the spatial
grid size. The time for the onset of collapse appears to tend to a finite
limit when the grid size approaches zero, but the height of a peak in
the collapsed region seems to grow without bound, faster than (grid
size)$^{-1}$. This is consistent with the expectation that density
fluctuations become divergent in the continuum limit.

\section{Growth-Collapse Cycles}

We have ignored gain-loss mechanisms so far, because we wanted to understand
various effects one at a time. Now we shall take them into account, and make
the problem more realistic.

In the experimental situation, the condensate in the trap can exchange atoms
with an uncondensed ``thermal cloud'', which contains far more atoms than
the condensate. Equilibrium is reached when the chemical potentials
equalize. The equilibrium fraction of condensate to ``cloud'' atoms is
determined by the temperature. The kinetic equations governing the approach
to equilibrium are quite intricate, and a subject of ongoing research \cite
{kinetics}. We shall use a phenomenological gain equation based on a fit to
data \cite{growth}:

\begin{equation}
\frac{dN}{dt}=c_{0}N\left[ 1-\left( \frac{N}{N_{\text{eq}}}\right) ^{0.4}%
\right]
\end{equation}
where $N_{\text{eq}}$ is the equilibrium number of atoms in the condensate,
with 
\begin{equation}
c_{0}=f\gamma=f\sqrt{8}\zeta(3/2)\gamma_{\text{el}}(15G_{eq})^{0.4}\frac{\hbar\omega}{k_BT}
\label{equilnumber}
\end{equation}
where $\gamma_{\text{el}}$ is the elastic scattering rate, which
is approximately equal to 1 s$^{-1}$ for the conditions of the
experiment of Ref.\cite{Bradley} and $\zeta(3/2) \approx 2.612$. $f$
is a phenomenological factor to reconcile the above theoretical
prediction for $\gamma$ with the experimental results for the  
repulsive case \cite{growth}. In our runs we used $f = 5.75$ and
$T=300$ nK, while $G_{\text{eq}}$ was taken to be equal to the
critical value of $G$, i.e. about $0.574$. Assuming
$N/N_{\text{eq}}\ll 1$, we ignored the second term in the bracket.
The main loss mechanisms are two- and three-body collisions\cite{Dodd}
described by the following equation:

\begin{eqnarray}
\frac{dn}{dt} &=&-c_{1}n^{3}-c_{2}n^{2}  \nonumber \\
c_{1} &=&2.6\times 10^{-28}\text{ cm}^{6}\text{/s}  \nonumber \\
c_{2} &=&1.2\times 10^{-14}\text{ cm}^{3}\text{/s}
\end{eqnarray}
where $n$ is the particle density in the condensate. Adding the gain and
loss terms, we have an extended NLSE, which reads, in units such that $%
d_{0}=2\omega ^{-1}=1$,

\begin{equation}
i\frac{\partial u}{\partial t}=\left( -\frac{\partial ^{2}}{\partial r^{2}}%
+r^{2}-\frac{2G|u|^{2}}{r^{2}}\right) u+i\left( \gamma _{0}-\gamma _{1}\frac{%
G^{2}|u|^{4}}{r^{4}}-\gamma _{2}\frac{G|u|^{2}}{r^{2}}\right) u
\end{equation}
where 
\begin{eqnarray}
\gamma _{0} &=&2.6\times 10^{-3}  \nonumber \\
\gamma _{1} &=&8.3\times 10^{-6}  \nonumber \\
\gamma _{2} &=&7.2\times 10^{-5}
\end{eqnarray}
In Ref.\cite{Kagan2}, the value of $\gamma _{1}$ is taken to be of order $%
10^{-3}$, which differs considerably from ours.

With the new equation, we find that the initial conditions become quite
irrelevant. The number of particles quickly settle into a steady-state
oscillation whose peak value defines the critical particle
number in the new context. This is shown in Fig.9, with: 
\begin{equation}
N_{c}\approx 1260
\end{equation}
The mean number of atoms is approximately 735, and the period of the
oscillation is 
\begin{equation}
\tau \approx \frac{917}{\omega }=1.01\text{ s}
\end{equation}
The long time scale of the oscillations explains why they are
insensitive to initial conditions. We have extended the calculations
to the longest evolution time practicable, and found that, with a
constant supply of particles, $N$ oscillates in a steady state for at
least 1.44$\times 10^{4}$ trap cycles, or 100 seconds. These results
are comparable with those obtained by Hulet et al. \cite{Hulet2},
except that, in contrast to our case, $N$ collapses to zero in that
calculation.

\section*{Acknowledgment}
\noindent
This work is support in part by a DOE cooperative
agreement DE-FC02-94ER40818.

\newpage

\section*{\protect\bigskip\large Figures}

\begin{figure}
$$
\BoxedEPSF{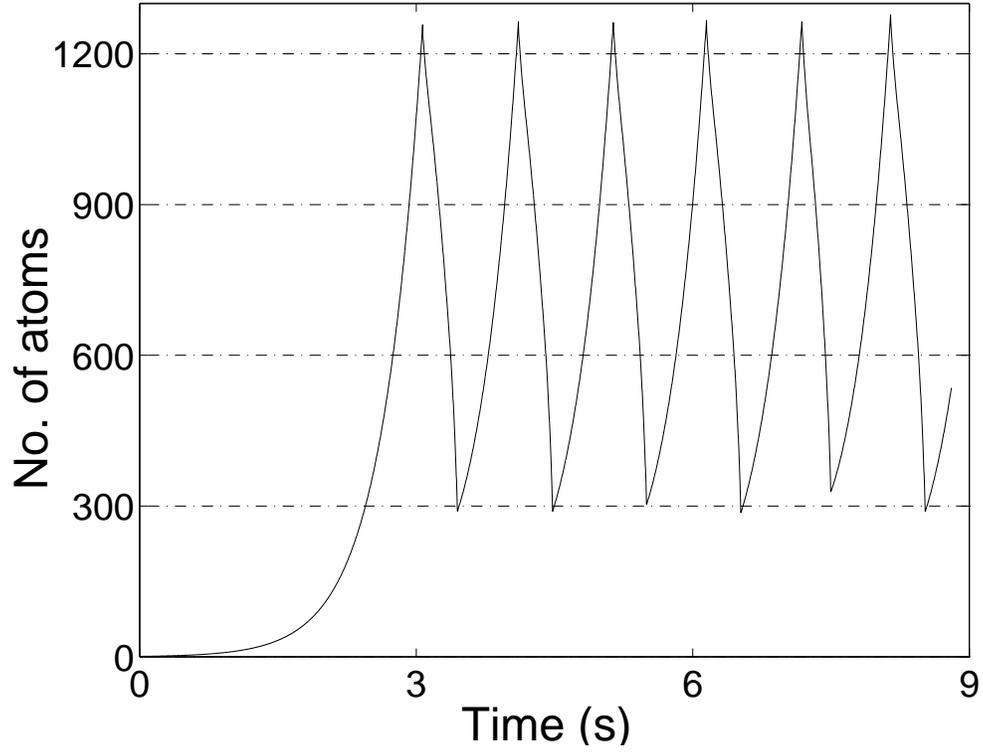 scaled 700} 
$$
\caption{Steady-state oscillation of the number of particles, when the
 condensate is fed by a ``thermal cloud'' and suffers loss through two- and
 three-body collisions. The peaks give $N_{c}\approx 1260$. The period
 of 1.01 s is much greater than the trap period of 7 ms. For this reason,
 initial conditions are irrelevant.}
\label{AEKHFig9}
\end{figure}

\end{document}